%% file: main.tex
\documentclass{article}
\usepackage{spconf,amsmath,graphicx, amssymb, cite, caption}
\usepackage[breaklinks=true]{hyperref}

\title{Neural synthesis of footsteps sound effects \\ with generative adversarial networks}
\name{Marco Comunità, Huy Phan, Joshua D. Reiss}
\address{Centre for Digital Music, Queen Mary University of London, UK}
\begin{document}
\ninept
\maketitle

\begin{abstract}
Footsteps are among the most ubiquitous sound effects in multimedia applications. There is substantial research into understanding the acoustic features and developing synthesis models for footstep sound effects. In this paper, we present a first attempt at adopting neural synthesis for this task. We implemented two GAN-based architectures and compared the results with real recordings as well as six traditional sound synthesis methods. Our architectures reached realism scores as high as recorded samples, showing encouraging results for the task at hand.
\end{abstract}
%

\vspace{-4pt}
\section{Introduction}\label{sec:intro}
\vspace{-4pt}
When sound designers are given the task of creating sound effects for movies, video games or radio shows, they have essentially two options: pre-recorded samples or procedural audio. In the first case, they usually rely on large libraries of high quality audio recordings, from which they have to select, edit and mix samples for each event they need to sonify. For a realistic result, especially in video games where the same actions are repeated many times, several samples are selected for every event and randomised during action. This creates challenges in terms of memory requirements, assets management and implementation time. 

Alternatively, procedural audio \cite{farnell2010designing} aims to synthesise sound effects in real time, based on a set of input parameters. In the context of video games, these parameters might come from the specific interaction of a character with the environment. This approach presents challenges in terms of development of procedural models which can synthesise high quality and realistic audio, as well as finding the right parameters values for each sound event.

Footsteps sounds are a typical example of the challenges of sound design. It is an omnipresent sound, which is generally fairly repetitive, but on which the human ear is being constantly trained. In fact, it is possible for a subject to identify from recorded or synthesised footsteps, things like: gender \cite{li1991perception}, emotions \cite{giordano2006walking}, posture \cite{pastore2008auditory}, identity \cite{makela2003use}, ground materials \cite{nordahl2010sound}, and type of locomotion \cite{bresin2003experiments}.

Thus, it is not surprising that researchers put substantial efforts into understanding the acoustic features \cite{turchet2016your} and developing synthesis models of footsteps sounds. Cook \cite{cook2002modeling} made a first attempt at synthesising footsteps on different surfaces, based on his previous work on physically-informed stochastic models (PhISM) \cite{cook1997physically}. Another physically-informed model - based on using a stochastic controller to drive sums of microscopic impacts - was proposed by Fontana and Bresin \cite{fontana2003physics}. DeWitt and Bresin \cite{dewitt2007sound} proposed a model that included the user's emotion parameter. Farnell \cite{farnell2007marching} instead, developed a procedural model by studying the characteristics of locomotion in primates. In \cite{turchet2016footstep} Turchet developed physical and physically-inspired models coupled with additive synthesis and signals multiplication.

To this day, there has not yet been an attempt at exploring the use of neural networks for the synthesis of footsteps sounds although there is substantial literature exploring neural synthesis of broadband impulsive sounds, such as drums samples, which have some similarities to footsteps. One of the first attempts was in \cite{donahue2018adversarial}, where Donahue \emph{et al.} developed WaveGAN - a generative adversarial network for unconditional audio synthesis. Another example of neural synthesis of drums is \cite{nistal2020drumgan}, where the authors used a Progressive Growing GAN. Variational autoencoders \cite{aouameur2019neural} and U-Nets \cite{ramires2020neural} have also been used for the same task. But, the application of recent developments to neural synthesis of sound effects is yet to be explored. We could only find one other work \cite{barahona2020synthesising}, related to the present study, where the authors focused on synthesis of knocking sounds with emotional content using a conditional WaveGAN.

In this paper we describe the first attempt at neural synthesis of footsteps sound effects. In Section \ref{sec:architecture}, we  propose a hybrid architecture  that improves the quality and better approximates the real data distribution, with respect to a standard conditional WaveGAN. Objective evaluation of this architecture is provided in Section \ref{sec:obj_eval}. Section \ref{sec:Subj_eval} reports on the first listening test that compares ``traditional'' and neural synthesis models on the task at hand. Discussion and conclusions are given in Section \ref{sec:discussion}.

\vspace{-4pt}
\section{Architecture}\label{sec:architecture}
\vspace{-4pt}

\vspace{-4pt}
\subsection{Data}\label{sec:data}
\vspace{-4pt}
To train our models we curated a small dataset (81 samples) using free footsteps sounds available on the Zapsplat website\footnote{\href{https://www.zapsplat.com/sound-effect-packs/footsteps-in-high-heels}{https://www.zapsplat.com/sound-effect-packs/footsteps-in-high-heels}}. These are high quality samples, recorded by Foley artists using a single type of shoes (women, high heels) on seven surfaces (carpet, deck, metal, pavement, rug, wood and wood internal). The samples were converted to WAV file format, resampled at 16kHz, time aligned and normalised to -6dBFS.

\vspace{-4pt}
\subsection{Generator}\label{sec:generator}
\begin{figure}
	\centering
	\includegraphics[width=.7\linewidth]{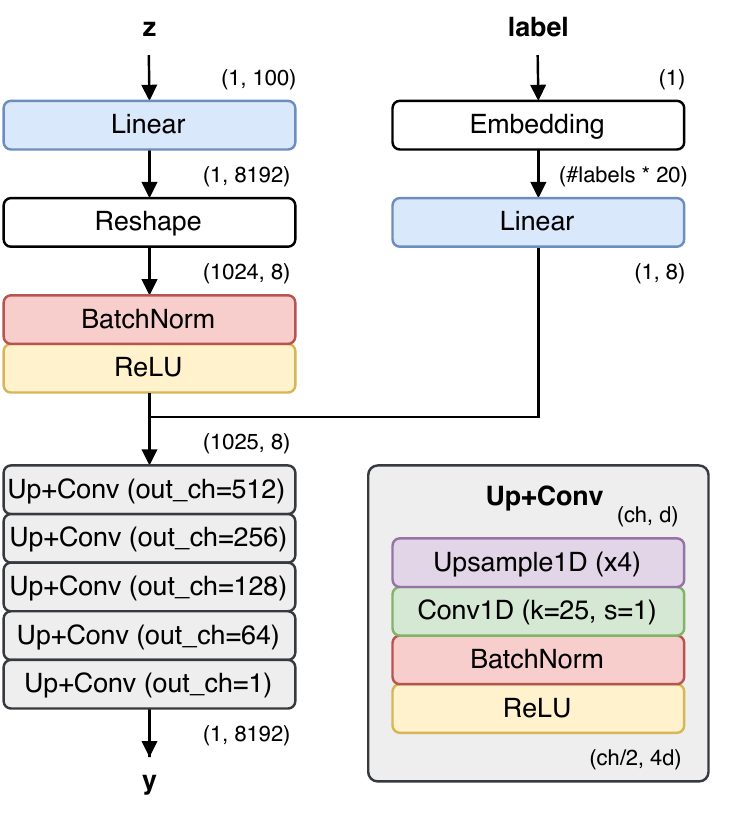}
	\caption{WaveGAN generator.}
    \label{fig:wave_gen}
\end{figure}

\vspace{-4pt}
The original WaveGAN generator \cite{donahue2018adversarial} was designed to synthesise 16384 samples ($\sim$1s at 16kHz sampling frequency); and afterwards extended to 32768 or 65536 samples\footnote{\href{https://github.com/chrisdonahue/wavegan}{https://github.com/chrisdonahue/wavegan}} ($\sim$2s and $\sim$4s at 16kHz). We adapted the architecture to synthesise 8192 samples, which is sufficient to capture all footsteps samples in our dataset. As shown in Fig.~\ref{fig:wave_gen}, the generator expands the latent variable $z$ to the final audio output size. After reshaping and concatenation with the conditioning label \cite{barahona2020synthesising, lee2018conditional}, the output is synthesised by passing the input through 5 upsampling 1-D convolutional layers. Upsampling is obtained by: zero-stuffing or nearest neighbour, linear or cubic interpolation. Zero-stuffing plus 1-D convolution is equivalent to using 1-D transposed convolution with stride equal to the upsampling rate. In the other cases, we split the operation into upsampling and 1-D convolution with stride of 1. Every convolutional layer uses a kernel size of 25, as in the original design. Differently from the original, we included batch normalisation layers after each convolution. With the exception of the last layer, the number of output channels is halved at each convolution layer. 

\vspace{-4pt}
\subsection{Discriminator}\label{sec:discriminator}
\vspace{-4pt}
Recent results in the field of neural speech synthesis \cite{kumar2019melgan, yamamoto2020parallel, jang2020universal, song2021improved, yang2020vocgan, kong2020hifi} have shown how GAN-based vocoders are capable of reaching state-of-the-art results in terms of mean opinion scores. You \emph{et al.} \cite{you2021gan} hypothesised that this success is not related to the specific design choices or training strategies; they identified it in the multi-resolution discriminating framework. In their work, the authors trained 6 different generators using the same discriminator (Hifi-GAN - \cite{kong2020hifi}) reaching very similar performance independently of the specific generator.

\begin{figure}
	\centering
	\includegraphics[width=.67\linewidth]{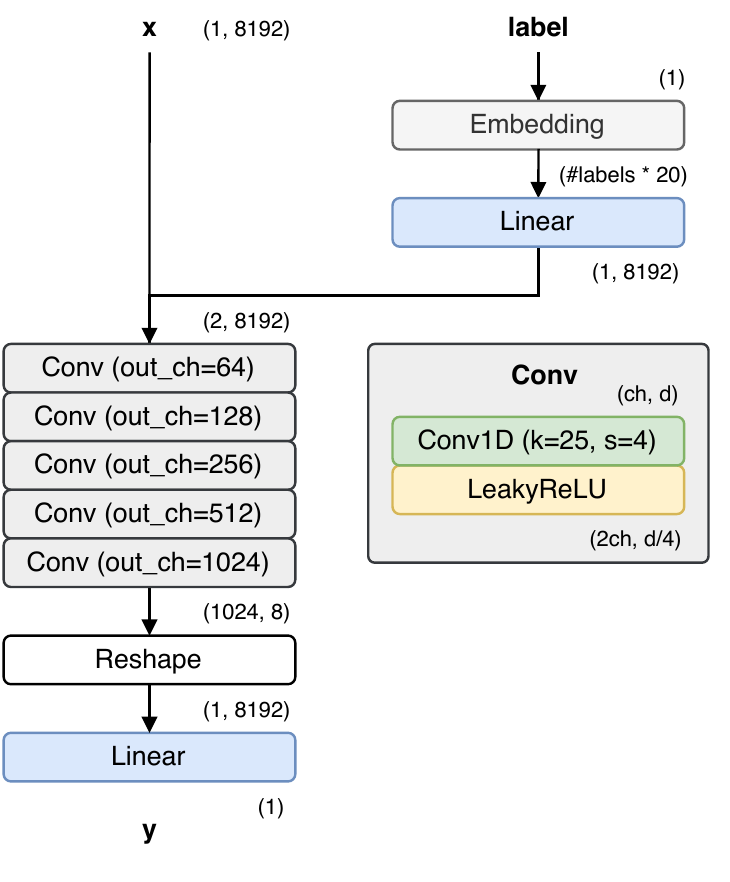}
	\vspace{-0.1cm}
	\caption{WaveGAN discriminator.}
    \label{fig:wave_dis}
\end{figure}

We experimented with a similar approach by implementing conditional versions of WaveGAN and HiFi-GAN discriminators. Our WaveGAN discriminator was again adapted to 8192 samples of the original architecture, based on 5 1-D convolutional layers with stride of 4 (see Fig.~\ref{fig:wave_dis}).
The HiFi-GAN discriminator is made of two separate discriminators (multi-scale and multi-period) each of which is made of several sub-discriminators that work with inputs of different resolutions. The multi-scale discriminator works on raw audio, $\times2$ average-pooled and $\times4$ average-pooled audio (i.e., a downsampled and smoothed version of the original signal). The multi-period discriminator works on ``equally spaced samples of an input audio; the space is given as period $p$''. The periods are set in the original model to 2, 3, 5, 7 and 11.

\vspace{-4pt}
\subsection{Loss and Training Procedure}\label{sec:training}
\vspace{-4pt}
We trained the two architectures - WaveGAN generator and discriminator (referred to as WaveGAN), WaveGAN generator and HiFi-GAN discriminator (referred to as HiFi-WaveGAN) - using different paradigms. 

In the first case we opted for a Wasserstein GAN with gradient penalty \cite{gulrajani2017improved} (WGAN-GP), which has been shown to improve training stability and help convergence towards a minimum which approximates better the real data distribution as well as the synthesised samples' quality. When training a WGAN-GP the discriminator's weights were updated several times for each update of the generator; we followed the standard approach of 5 to 1 updates ratio.

For HiFi-WaveGAN we followed the approach suggested in \cite{kong2020hifi}, opting for least squares GAN (LS-GAN) \cite{mao2017least}, where the binary cross-entropy terms of the original GAN \cite{goodfellow2014generative} are replaced with least squares losses. \cite{kong2020hifi} included additional losses for the generator; specifically, a mel-spectrogram loss and a feature matching loss. The mel-spectrogram loss measures the $\ell1$-distance between the mel-spectrogram of a synthesised and a ground truth waveforms. We discarded this term since, differently from HiFi-GAN, our generator was not designed and trained to synthesise waveforms from ground-truth spectrograms. However, we kept the feature matching loss, which measures the $\ell1$-distance of the features extracted at every level of each sub-discriminator, between real and generated samples. See \cite{kong2020hifi} for a more detailed description of each loss term.
The final losses for our HiFi-WaveGAN were:
\begin{align*}
    \mathcal{L}_{G} &= \mathcal{L}_{Adv}(G; D) + \lambda_{fm}\mathcal{L}_{FM}, \\
    \mathcal{L}_{D} &= \mathcal{L}_{Adv}(D; G),
\end{align*}
where $\mathcal{L}_{G}$ and $\mathcal{L}_{D}$ are the generator and discriminator total losses with $\mathcal{L}_{Adv}(G; D)$ and $\mathcal{L}_{Adv}(D; G)$ the adversarial loss terms for generator and discriminator and $\lambda_{fm}\mathcal{L}_{FM}$ the feature matching loss.
Both architectures were trained for 120k batches, with a batch size of 16 and a learning rate of 0.0001. WaveGAN used Adam while HiFi-WaveGAN used AdamW optimisers.

\vspace{-4pt}
\section{Objective Evaluation}\label{sec:obj_eval}
\vspace{-4pt}
\begin{figure*}[ht]
    \begin{minipage}{.33\linewidth}
      \centering
      \includegraphics[width=.7\linewidth]{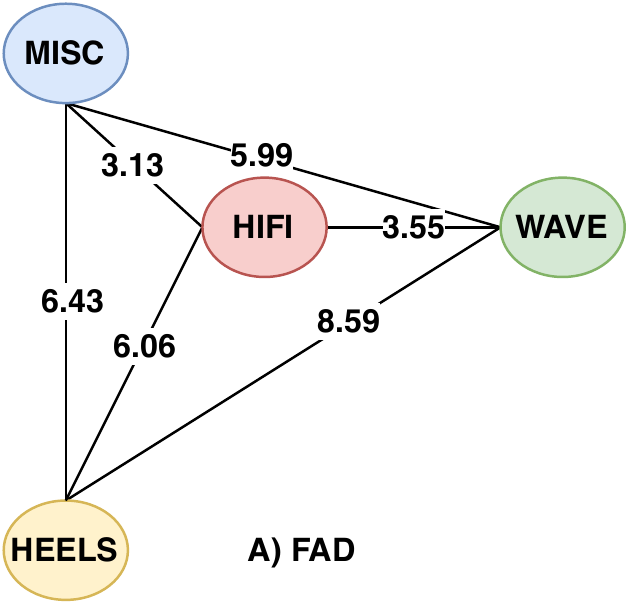}
    \end{minipage}%
    \begin{minipage}{.33\linewidth}
      \centering
      \includegraphics[width=.52\linewidth]{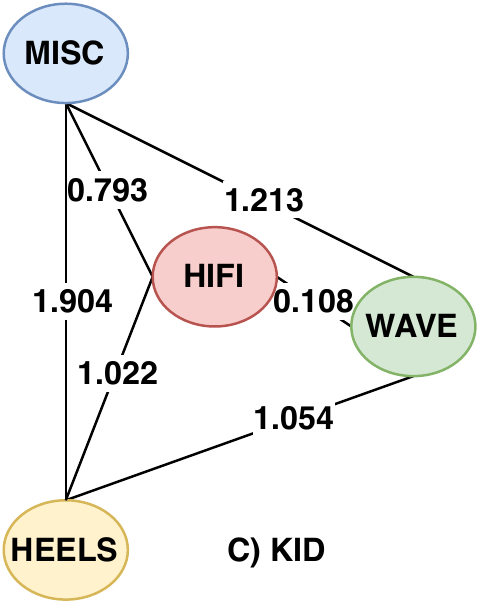}
    \end{minipage}
    \begin{minipage}{.33\linewidth}
      \centering
      \includegraphics[width=.67\linewidth]{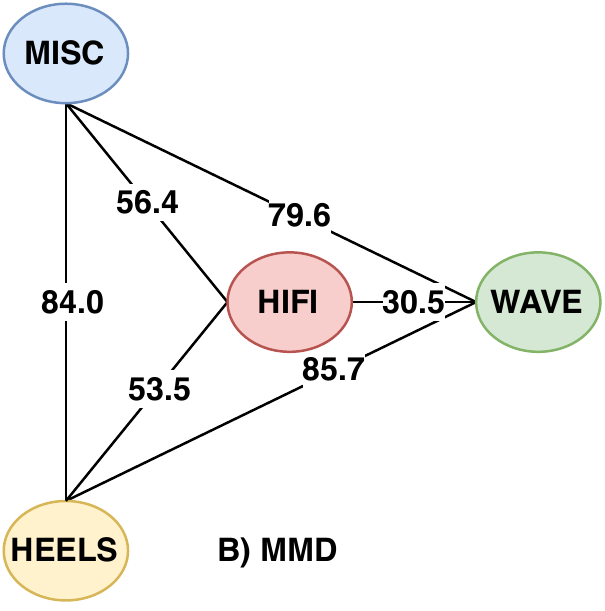}
    \end{minipage}
    \caption{Graphs representing Fr\'echet Audio Distance (A), Kernel Inception Distance (B) and Maximum Mean Discrepancy (C) for our models.}
    \label{fig:fad_kid_mmd}
\end{figure*}
There are no formalised methods to reliably evaluate the quality and diversity of synthesised audio, but there are several metrics which are commonly adopted to analyse and compare neural synthesis models. We followed a similar approach to \cite{nistal2020drumgan} for objective evaluation, where the authors relied on Inception Score (IS), Kernel Inception Distance (KID) and Fr\'echet Audio Distance (FAD). We also relied on the maximum mean discrepancy (MMD) \cite{gretton2012kernel} as a measure of similarity between real and synthesised samples, using the same formulation adopted in \cite{turian2021one}, and computed the MMD using the $\ell1$-distance between OpenL3 embeddings \cite{cramer2019look} (env, mel128, 512)\footnote{\href{https://github.com/torchopenl3/torchopenl3}{https://github.com/torchopenl3/torchopenl3}}.

\begin{figure*}[ht]
    \begin{minipage}{.33\linewidth}
      \centering
      \includegraphics[width=1\textwidth]{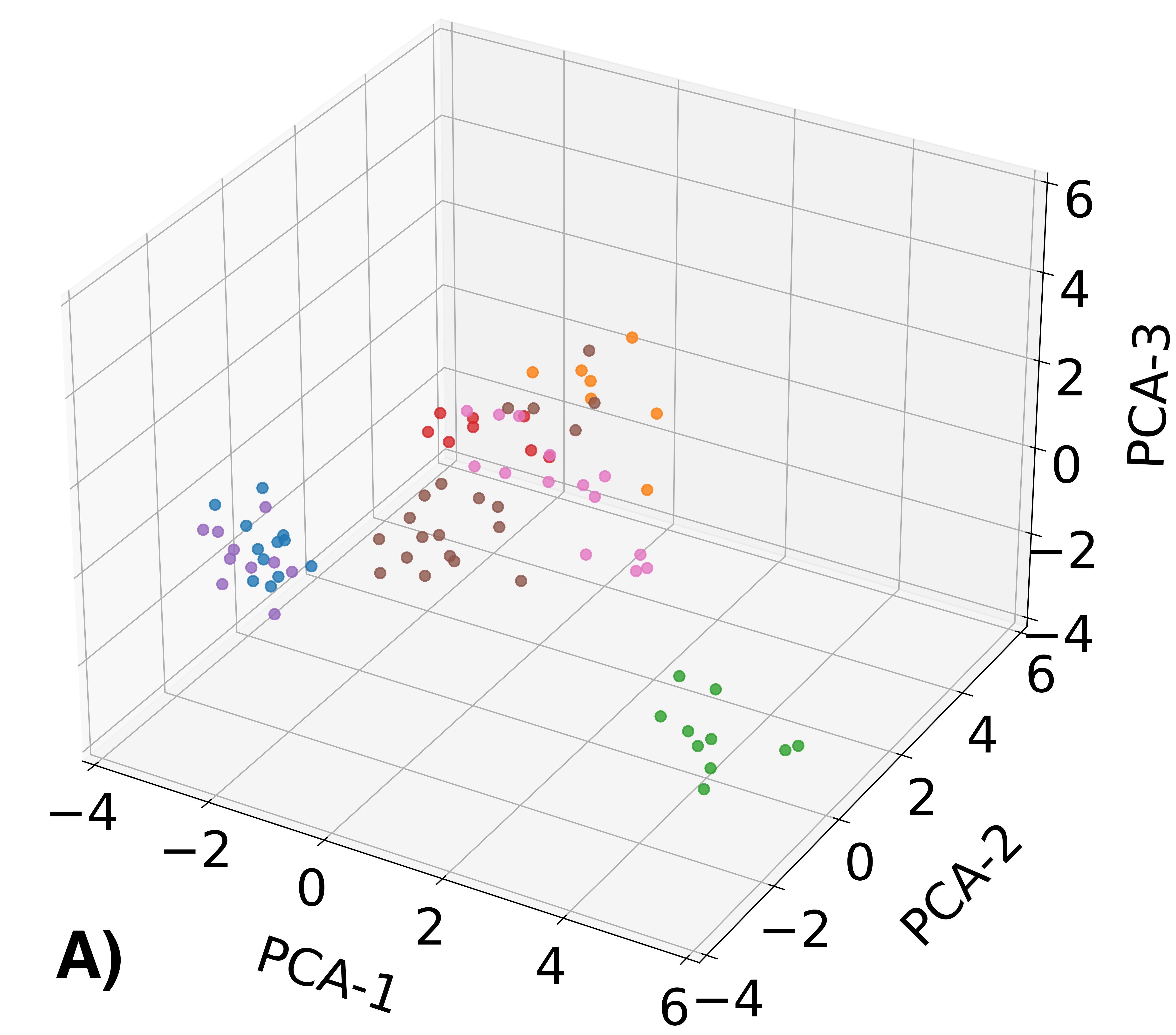}
    \end{minipage}%
    \begin{minipage}{.33\linewidth}
      \centering
      \includegraphics[width=1\textwidth]{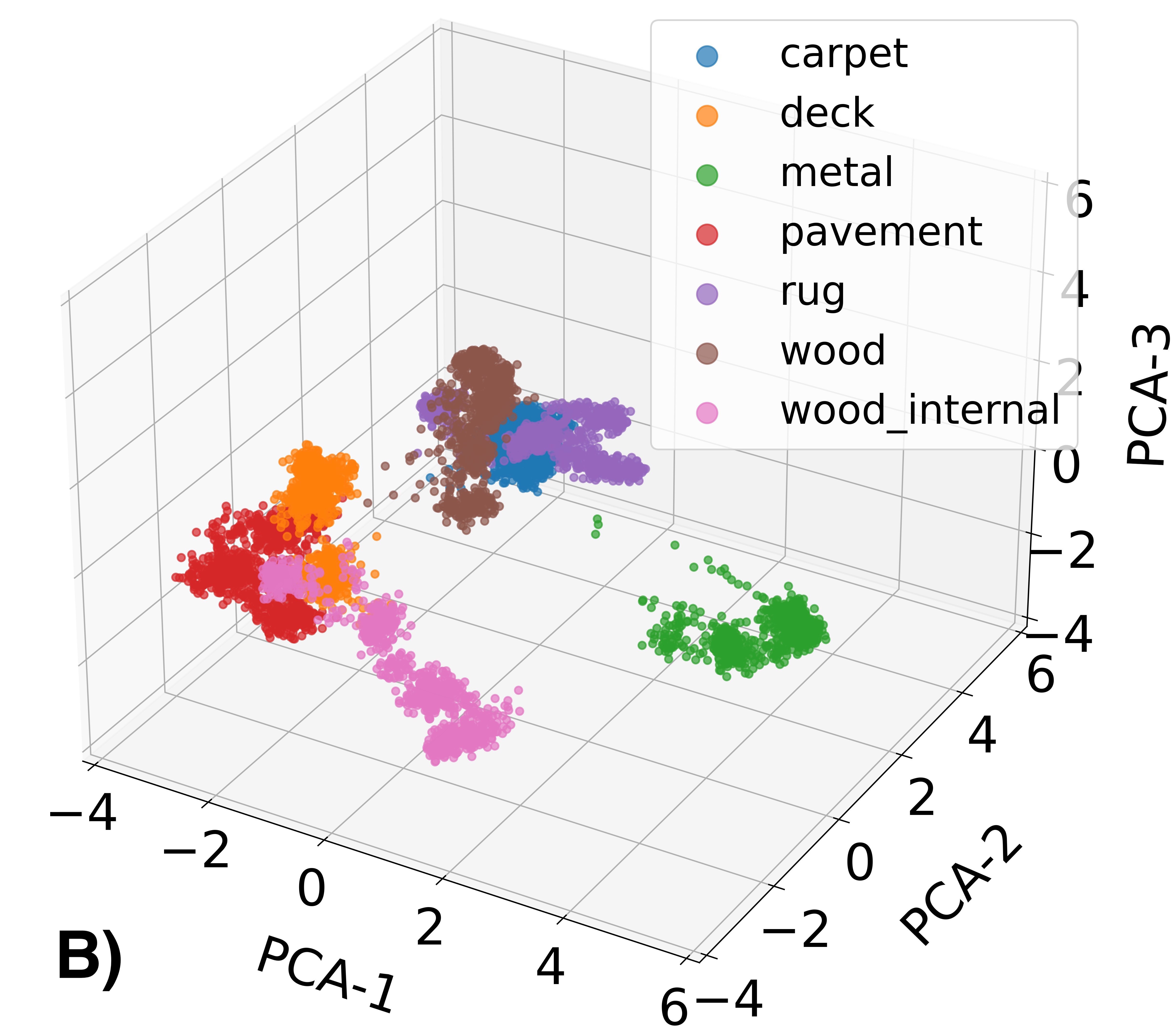}
    \end{minipage}
    \begin{minipage}{.33\linewidth}
      \centering
      \includegraphics[width=1\textwidth]{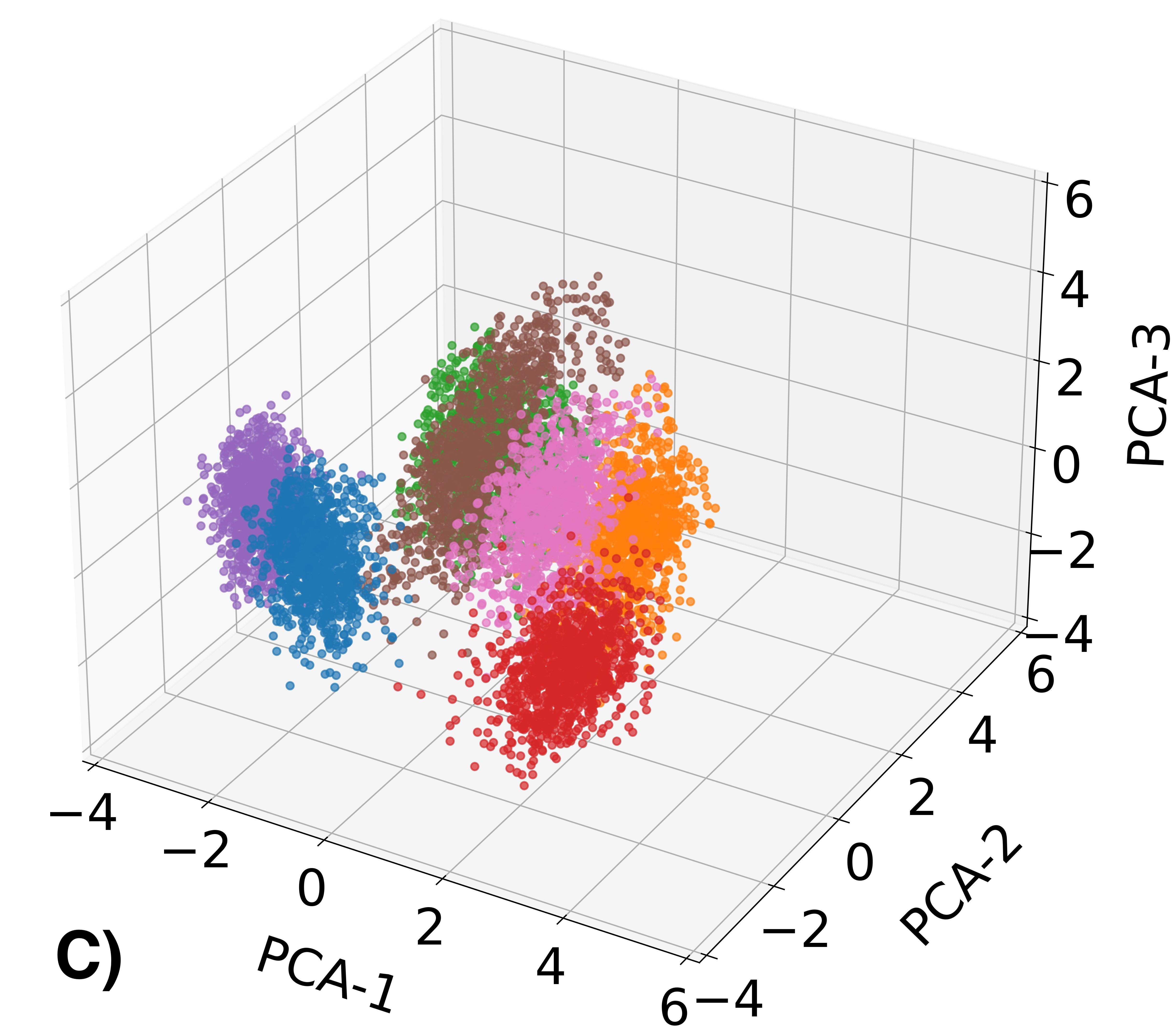}
    \end{minipage}
    \caption{Scatter plots of principal components analysis on OpenL3 embeddings for: A) Training Data, B) HiFi-WaveGAN and C) WaveGAN.}
    \label{fig:pca}
\end{figure*}

To analyse whether our models were capable of learning the real data distribution - and to verify that they were not affected by overfitting or mode collapse - we used principal components analysis (PCA) on the OpenL3 embeddings of real and synthesised samples. We generated 1000 samples for each class and plot the results of PCA in Fig.~\ref{fig:pca}. It can be seen that HiFi-WaveGAN reached a better approximation of the real data without collapsing into synthesising only samples that were seen during training. However, there is currently no way to establish a correlation between distance in the embedding space with distance in terms of human perception. It is therefore not possible to comment on the diversity of synthesised samples from these results.

Instead, we used IS, which gave us a way to measure diversity and semantic discriminability. This measure ranges from 1 to $n$ (with $n$ number of classes) and it is maximised for models which can generate samples for all possible classes and that are classified with high confidence. As classifier, we used the same Inception Net variant developed in \cite{nistal2020drumgan} and adapted it to our domain. We trained the model on a separate dataset (\textit{Zapsplat Misc}) obtained by scraping all the freely available footsteps samples on the Zapsplat website \footnote{\href{https://www.zapsplat.com}{https://www.zapsplat.com}} and organised them into 5 classes depending on the surface material (\textit{carpet/rug, deck/boardwalk, metal, pavement/concrete, wood/wood internal}). Likewise, we merged the training (\textit{Zapsplat Heels}) and generated data into the same 5 classes by grouping \textit{carpet/rug} and \textit{wood/wood internal} samples together. 

We trained the Inception Net for 100 epochs, reaching 86\% validation accuracy. The results are shown in Table~\ref{table:is}. It is interesting to notice how the IS for HiFi-WaveGAN is slightly higher then it is for the training data (\textit{Zapsplat Heels}). Assuming that the synthesised data cannot reach a higher semantic discriminability than the data it learned from, the score should be related to diversity. Which in turn might depend on two factors. First, we evaluated IS on a number of Hifi-WaveGAN samples orders of magnitudes greater than the training data (3500 versus 81). Excluding the case of mode collapse, this will inherently carry higher diversity. And second, Hifi-WaveGAN is actually able to synthesise samples not seen during training, and could therefore lead to a more diverse set. However, it is not possible to comment on the ``perceptual'' diversity of the generated samples based on IS only.

Fig.~\ref{fig:fad_kid_mmd} shows the other metrics we adopted to compare real (\textit{Misc} and \textit{Heels}) and synthesised (\textit{HiFi} and \textit{Wave}) data. For clarity, we represent FAD, KID and MMD on 3 graphs, where the distance between nodes is roughly proportional to each metric values (written on the edges). The 3 metrics, which are based on comparing embeddings distributions (VGG-ish for FAD, Inception for KID and OpenL3 for MMD), all depict a similar picture where HiFi-WaveGAN seems to better approximate the training data. FAD and KID also place HiFi-WaveGAN samples nearer to the \textit{Misc} dataset, again suggesting that the model is capable of synthesising samples with greater diversity than the training data. FAD, correlating with human judgement, is also a measure of the perceived quality of individual sounds (the lower the FAD, the higher the quality). In our case, HiFi-WaveGAN also seems to score higher in terms of quality.

\input{tables/is}

\vspace{-4pt}
\section{Subjective Evaluation}\label{sec:Subj_eval}
\vspace{-4pt}
For the subjective evaluation we followed a similar paradigm to \cite{moffat2018perceptual}. The Web Audio Evaluation Tool \cite{jillings2015web} was adopted to run an audio perceptual evaluation (APE) \cite{de2014ape}; a multi-stimulus test in which a participant was presented with a series of samples to be compared and rated on a continuous scale from 0 to 1. On this scale, 4 reference values (very unrealistic, somewhat unrealistic, somewhat realistic, very realistic) were given at the 0, 0.33, 0.66 and 1 points, respectively. Eight synthesis methods were compared to real recordings:
\begin{enumerate}
    \item Procedural model 1 (PM1) by Fontana and Bresin \cite{fontana2003physics}
    \vspace{-2pt}
    \item Procedural model 2 (PM2) by Farnell \cite{farnell2007marching} \footnote{\href{http://aspress.co.uk/sd/practical26.html}{http://aspress.co.uk/sd/practical26.html}}
    \vspace{-2pt}
    \item Procedural model 3 (PM3) from Nemisindo \cite{bahadoran2018fxive} \footnote{\href{https://nemisindo.com/}{https://nemisindo.com/}}
    \vspace{-2pt}
    \item Sinusoidal plus stochastic (SPS) by Amatriain \emph{et al.} \cite{amatriain2002spectral} \footnote{\href{https://www.dafx.de/DAFX_Book_Page_2nd_edition/chapter10.html}{https://www.dafx.de/DAFX\_Book\_Page\_2nd\_edition/chapter10.html}}
    \vspace{-2pt}
    \item Statistical modelling (STAT) by McDermott \emph{et al.}\cite{mcdermott2011sound} \footnote{\href{https://mcdermottlab.mit.edu/downloads.html}{https://mcdermottlab.mit.edu/downloads.html}}
    \vspace{-2pt}
    \item Additive synthesis (ADD) by Verron \emph{et al.}\cite{verron20093} \footnote{\href{http://www.charlesverron.com/spad.html}{http://www.charlesverron.com/spad.html}}
    \vspace{-2pt}
    \item WaveGAN (WAVE)
    \vspace{-2pt}
    \item HiFi-WaveGAN (HIFI)
\end{enumerate}
To present a more realistic and reliable scenario we prepared 10 s long walks by concatenating single samples. We started from real recordings and chose - through informal listening - the time interval between samples that gave the most realistic result. The same pace was then replicated for all the other synthesis methods. We prepared a total of 10 series of 9 walks (1 for each synthesis method plus the real recordings); and presented each participant with 5 of these series to be able to compare many different conditions (i.e., shoe types and surface materials) while keeping the test short.

\vspace{-3pt}
\subsection{Results}\label{sec:results}
\vspace{-3pt}
A total of 19 participants took part in the online test\footnote{\href{http://webprojects.eecs.qmul.ac.uk/mc309/FootEval/test.html?url=tests/ape_footsteps.xml}{http://webprojects.eecs.qmul.ac.uk/mc309/FootEval/ test.html?url=tests/ape\_footsteps.xml}}. Of these, 10 identified as male, 6 as female and 3 preferred not to indicate their gender. 3 participants were excluded since they had no previous experience with critical listening tests. Of the remaining participants, all but 1 had experience as: musicians (15 out of 16, $\mu=9.7$, $\sigma=7.6$ and max~$=23$ years), sound engineers (11 out of 16, $\mu=4.6$, $\sigma=7.6$, max~$=15$ years) and sound designers (7 out of 16, $\mu=1.7$, $\sigma=2.9$, max~$=10$ years). We also enquired about the headphone models and verified that all participants used good quality devices.

Two criteria were adopted to judge the reliability of each rating. We used one of the procedural models (PM1) as an anchor, and excluded all cases where this model was rated above 0.1. Also, we considered real recordings to be the reference, and excluded all cases where they were rated below 0.5.

\begin{figure}
    \vspace{0.2cm}
	\centering
	\includegraphics[width=0.95\linewidth]{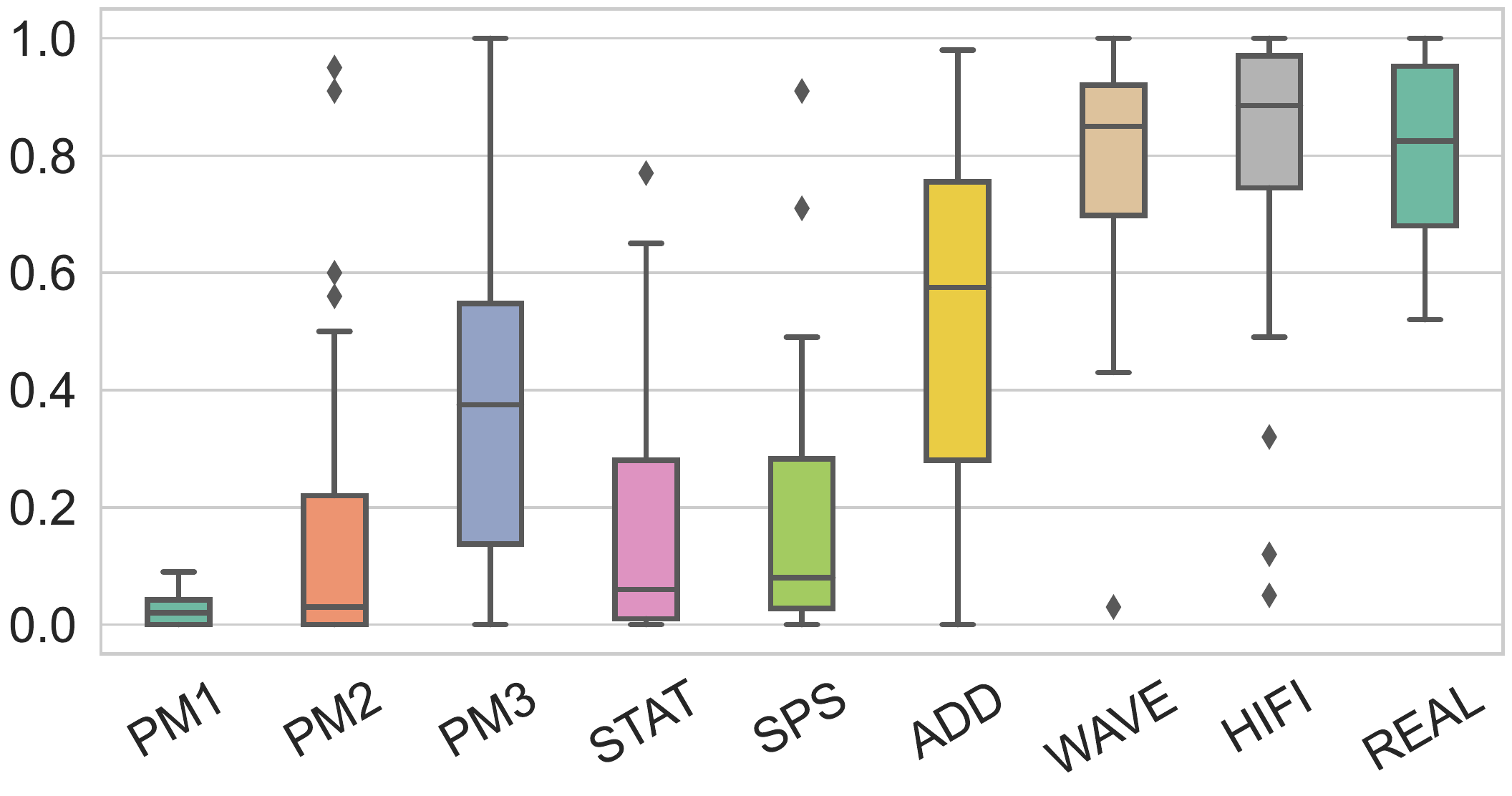}
	\caption{Results of the subjective evaluation.}
    \label{fig:bp_ratings}
\end{figure}

Final results are shown in Fig.~\ref{fig:bp_ratings}. Together with the anchor (PM1), Farnell's model (PM2), as well as statistical and sinusoidal modelling, are the lowest rated. This result is not surprising since Farnell's procedural model is a fairly basic implementation, which lacks the necessary output quality. Statistical modelling is usually adopted for texture synthesis, and also in those situations, it shows good results only for very specific cases (see \cite{mcdermott2011sound, moffat2018perceptual}). Also, sinusoidal modelling is more suitable for harmonic sounds (e.g. musical instruments) where the broadband, noisy components play a minor role in the overall spectrum. Nemisindo (PM3), being a more recent and advanced procedural model, scores higher than the other two; and additive synthesis, by synthesising sounds as a sum of five core elements (modal impact, noisy impact, chirped impact, band-limited noise, equalised noise), is rated even higher. Both WaveGAN and HiFi-WaveGAN score as high as the reference. This result shows that the two neural synthesis models manage to capture all the important details of the training data, and allow the generated samples to reach synthesis quality and realism comparable to recorded audio.

\vspace{-4pt}
\section{Discussion}\label{sec:discussion}
\vspace{-4pt}
We presented a first attempt at neural synthesis of footsteps - one of the most common and challenging sound effects in sound design. In this work, two GANs architectures were implemented: a standard conditional WaveGAN, and a hybrid consisting of a conditional WaveGAN generator and a conditional HiFi-GAN discriminator. The hybrid architecture improves the audio quality of generated samples and better approximates the training data distribution. Differently from what is commonly suggested in the literature, upsampling with zero-stuffing gave us better results than nearest neighbour interpolation. The same is true for linear or cubic interpolation.

Both objective and subjective evaluation of results were conducted. It is not common for neural synthesis methods to be compared with ``traditional'' synthesis algorithms in a listening test, which makes it impossible to establish whether a neural synthesis method can actually reach state-of-the-art results. In this work, we compared the two architectures with 6 other methods as well as recorded samples. 
The two architectures reached ``realism'' ratings as high as real sounds. Although, from informal listening tests, we would have expected a greater difference in the ratings. In fact, samples synthesised by WaveGAN are affected by a perceivable amount of background noise and distortion, while the opposite is true for HiFi-WaveGAN \footnote{\href{https://mcomunita.github.io/hifi-wavegan-footsteps_page/}{https://mcomunita.github.io/hifi-wavegan-footsteps\_page/}}.

This opens questions about the definition of realism of sound effects, to what extent audio quality correlates with perceived realism, and what other aspects play a significant role in such judgements. Following work will focus on increasing the degrees of control and diversity of synthesised samples, while retaining the audio quality.

\vspace{-4pt}
\section{Acknowledgements}\label{sec:acknowledgement}
\vspace{-4pt}
Funded by UKRI and EPSRC as part of the ``UKRI CDT in Artificial Intelligence and Music'', under grant EP/S022694/1.


\bibliographystyle{IEEEbib}
\bibliography{refs}

\end{document}

%% file: tables/is.tex
\begingroup
\setlength{\tabcolsep}{2pt} 
\renewcommand{\arraystretch}{1} 
\begin{table}
    \vspace{0.2cm}
    \small
    \centering
    \setlength{\tabcolsep}{4pt}
    \begin{tabular}{|c|c|}
        \hline
        \textbf{Dataset}
            & \textbf{IS} \\
        \hline
        Zapsplat Misc
            & 4.139 \\
        Zapsplat Heels
            & 3.221 \\
        HiFi-WaveGAN
            & 3.411 \\
        WaveGAN
            & 3.232 \\
        \hline
    \end{tabular}
    \caption{Inception Score for training and generated data.}
    \label{table:is}
\end{table}
\endgroup